\documentclass[pre,twocolumn, groupedaddress, showkeys,showpacs]{revtex4}

\usepackage{graphicx}
\begin{document}


\title{Particle diode: Rectification of interacting Brownian ratchets}



\author{Bao-quan  Ai$^{1}$} 
\author{Ya-feng He$^{2}$}
 \author{Wei-rong Zhong$^{3}$}

\affiliation{$^{1}$Laboratory of Quantum Information Technology, Institute for Condensed Matter Physics and School of Physics and Telecommunication
Engineering, South China Normal University, 510006 Guangzhou, China\\
 $^{2}$College of Physics Science and
Technology, Hebei University, 071002 Baoding, China\\
$^{3}$Department of Physics, College of Science and Engineering,
Jinan University, 510632 Guangzhou, China}


\date{\today}
\begin{abstract}
\indent Transport of Brownian particles interacting with each other via the Morse potential is
investigated in the presence of an ac driving force applied locally at one end of the chain.
By using numerical simulations, we find that the system can behave as a particle
 diode for both overdamped and underdamped cases. For low frequencies, the transport from the free end to the
 ac acting end is prohibited, while the transport from the ac acting end to the free
 end is permitted. However, the polarity of the particle diode will reverse for medium frequencies.
There exists an optimal value of the well depth of the interaction potential at which the
average velocity takes its maximum.  The average velocity $\upsilon$ decreases
monotonically with the system size $N$ by a power law $\upsilon \propto N^{-1}$.
\end{abstract}

\pacs{05. 40. Fb, 02. 50. Ey, 05. 40. -a}



\maketitle
\section {Introduction}
\indent Brownian motors \cite{a1}, rectifying nonequilibrium random walks on asymmetric potentials,
have recently been proposed for a variety of applications, including in biological systems\cite{a2},
as well as their potential technological applications ranging from classical non-equilibrium models\cite{a3}
to quantum systems\cite{a4}.  Ratchets have been proposed to model the unidirectional motion driven by
zero-mean non-equilibrium fluctuations. Broadly speaking, ratchet devices fall into three categories depending
 on how the applied perturbation couples to the substrate asymmetry:  rocking ratchets\cite{a5},
 flashing ratchets\cite{a6}, and correlation ratchets \cite{a7}. Additionally, entropic ratchets,
 in which Brownian particles move in a confined structure, instead of a potential, were also extensively studied \cite{a8}.

\indent In many physical situations, one deals with not a single Brownian particle, but rather the with arrays of a finite number of
interacting particles, which are usually in a periodic potential and acted upon by some external force.
There has been a lot of interest in recent years in the dynamics of interacting Brownian particles \cite{a9,a10,a11,a12,a13,a14}.
The reason for this interest is twofold \cite{a15}. First, experiments have provided a
wealth of information about the motion of individual colloidal particles.
A system of interacting Brownian particles is the simplest model of a colloidal suspension.
Second, interacting Brownian particles constitute the simplest model system, on which one can
test techniques and approximations of nonequilibrium statistical mechanics. Many-particle systems may exhibit some features not
found in the single-particle counterparts, such as phase transitions, spontaneous ratchet effects, and negative mobility \cite{a9,a10,a11,a12,a13,a14}.
The interacting Brownian ratchets been proposed for a variety of applications, including molecular motors \cite{a2}, friction \cite{a16},
diffusion of dimers on surfaces \cite{a17}, diffusion of colloidal particles \cite{a18}, DNA translocation through a nanopore \cite{a19}, charge density waves \cite{a20},
 and arrays of Josephson junctions \cite{a21}.

 \indent In the present work, we study the transport of Brownian  particles interacting with each other via the Morse potential by applying an ac driving  force
  at one end of the chain. From the numerical simulations, we find that the system can behave as a particle diode for both overdamped and underdamped
  cases, the transport is permitted when the ac driving force is applied at one end of
  the chain, while the transport is prohibited when the force acts on the other end of the chain.

\section{Model and Methods}
\begin{figure}[htbp]
  \begin{center}\includegraphics[width=7cm]{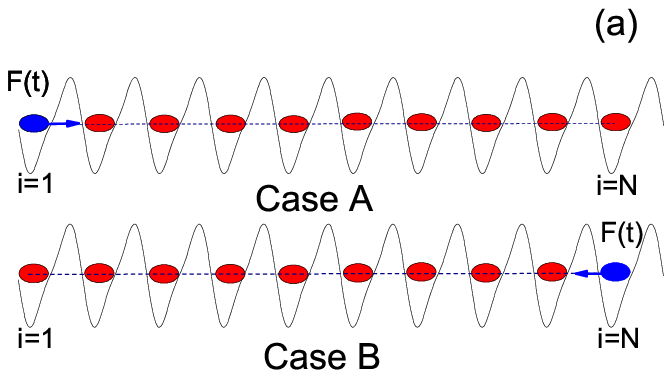}
  \includegraphics[width=7cm]{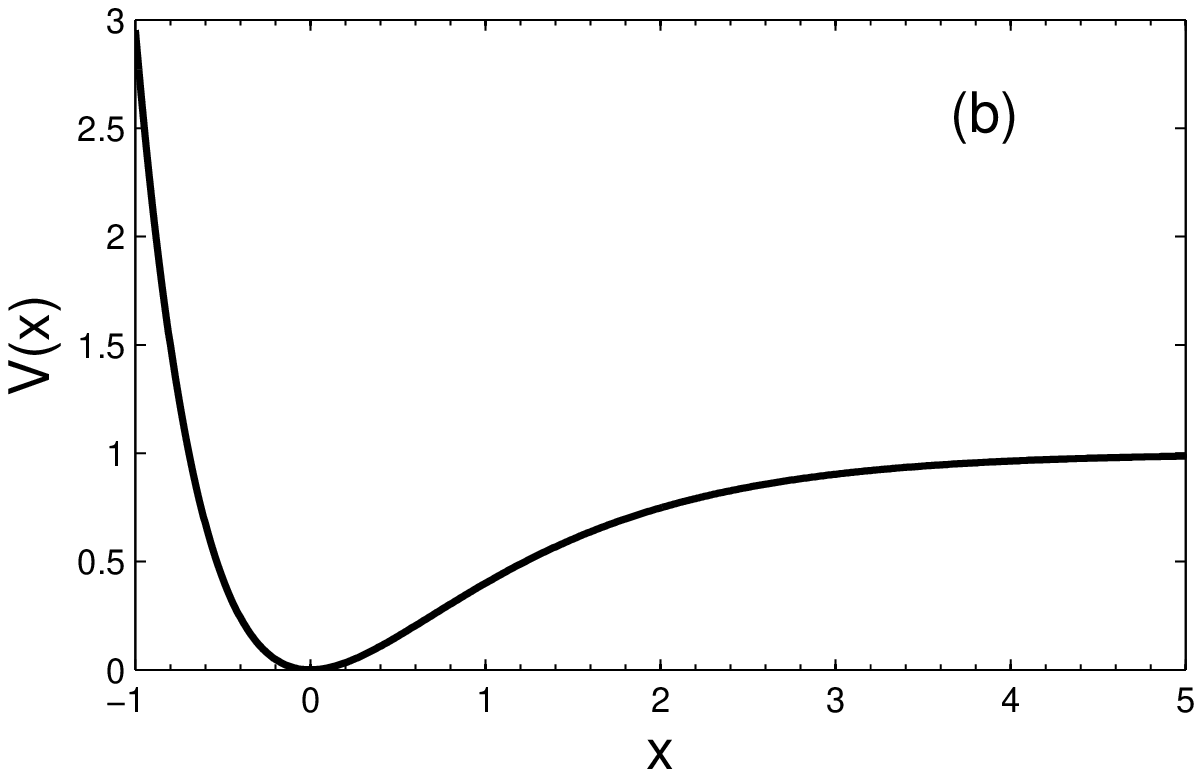}
  \caption{(Color online) (a) Schematic illustration of the model. Case A ($q=1$): the driving force acting on the left end.
  Case B ($q=N$): the driving force acting on the right end. The on-site potential $U(x)$ is asymmetric [shown in Eq. (7)].
   The ac driving force $F(t)=A_{0}\sin(\omega t)$. $U_{0}=1.0$ and $\Delta=1.0$.
   Note that it is possible that two particles share a site or that the chain is torn apart.
    (b) The interaction potential $V(x)$ is a Morse potential [shown in Eq.
   (8)]. The interaction potential is asymmetric. The parameters are
   $k=1.0$ and $a=1.0$.}\label{1}
\end{center}
\end{figure}

\indent We consider an open system of $N$ point-like Brownian particles interacting with each other via the Morse potential $V$,
with the substrate through the asymmetric periodic potential $U$. As to obtain the directed transport, a sustained
time-periodical force $F(t)$ is applied at one end of the chain.
The equations of motion for the underdamped case are
\begin{equation}\label{}
    m\ddot{x}_{1}=f(x_{1})-G(x_{2}-x_{1})-\gamma\dot{x_{1}}+\xi_{1}(t)+\delta_{q,1}F(t),
\end{equation}
\begin{equation}\label{}
    m\ddot{x}_{i}=f(x_{i})+G(x_{i}-x_{i-1})-G(x_{i+1}-x_{i})-\gamma\dot{x_{i}}+\xi_{i}(t),
\end{equation}
\begin{equation}\label{}
    m\ddot{x}_{N}=f(x_{N})+G(x_{N}-x_{N-1})-\gamma\dot{x}_{N}+\xi_{N}(t)+\delta_{q,N}F(t),
\end{equation}
where $x_{i}$ is the position of the $i$th particle, $m$ is the mass of the particles, and $N$ is total number of the particles.
$f=-\frac{\partial U}{\partial x}$ and $G=-\frac{\partial V}{\partial x}$. $\gamma$ is friction coefficient and the noise terms
$\xi_{i}(t)$ satisfy the fluctuation dissipation relations $\langle
\xi_{i}(t)\xi_{i}(t^{'})\rangle=2D\delta(t-t^{'})$. $D$ is the noise
intensity. The dot stands for the derivative with respect to time $t$. Where $q$ is a nature number. For $q=1$ and $q=N$, the ac driving force is applied at the left end
and the right end, respectively.

\indent When the inertia is negligible compared to the viscous damping,
Eqs. (1)-(3) can be rewritten for overdamped case,
\begin{equation}\label{}
\gamma\dot{x_{1}}=f(x_{1})-G(x_{2}-x_{1})+\xi_{1}(t)+\delta_{q,1}F(t),
\end{equation}
\begin{equation}\label{}
    \gamma\dot{x_{i}}=f(x_{i})+G(x_{i}-x_{i-1})-G(x_{i+1}-x_{i})+\xi_{i}(t),
\end{equation}
\begin{equation}\label{}
    \gamma\dot{x}_{N}=f(x_{N})+G(x_{N}-x_{N-1})+\xi_{N}(t)+\delta_{q,N}F(t).
\end{equation}

$U(x)$ is an asymmetrically periodic potential [shown in Fig. 1(a)]
\begin{equation}\label{}
    U(x)=-U_{0}[\sin(x)+\frac{\Delta}{4}\sin(2x)],
\end{equation}
where $U_{0}$ denotes the height of the potential and $\Delta$ is
its asymmetry parameter.

\indent The interaction potential between the nearest-neighbor particles is given by the Morse potential [shown in Fig. 1 (b)],
\begin{equation}\label{}
    V(x)=k[1-\exp(-a x)]^{2},
\end{equation}
where $k$ is well depth of the interaction potential and $a$
controls its width. The interaction potential is asymmetric and the
repulsive force is dominating.  $F(t)$ is a sustained time-periodical
force,
 \begin{equation}\label{}
    F(t)=A_{0}\sin(\omega t),
 \end{equation}
 where $A_{0}$ is the amplitude of the external force and $\omega$ is its
frequency.

\indent Here we mainly focus on the transport of the
driven particles. The average velocity $\upsilon_{i}$ of the $i$th particle is,
            \begin{equation}\label{}
            \upsilon_{i}(t)=\lim_{t\rightarrow\infty}\frac{\langle
            x_{i}(t)-x_{i}(t_{0})\rangle}{t-t_{0}},
            \end{equation}
where $t_{0}$ and $t$ are the initial and the end times for the
simulations, respectively. From Fig. 2 we can find that after the
system reaches a stationary state, $\upsilon_{i}$ is independent of
site position $i$. Therefore, we use the asymptotic velocity of the
particle $N/2+1$ to measure the transport of the system
$\upsilon=\upsilon_{N/2+1}$.

\begin{figure}[htbp]
  \begin{center}\includegraphics[width=7cm]{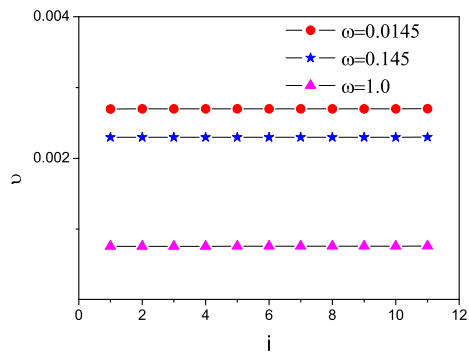}
  \caption{(Color online) The dependence of average velocity $\upsilon$ on the site $i$ for overdamped case.  It is found that each particle has the same average velocity.
   The other parameters are $k=1.0$, $A_{0}=2.0$,  $\Delta=1.0$, and $D=0.5$.}\label{1}
\end{center}
\end{figure}

\indent In our simulations, the equations of motion are integrated
by using the second order Stochastic Runge-Kutta algorithm\cite{b1}
with the small time step $\Delta t=10^{-3}$.  All quantities of
interest are averaged over 500 different realizations.  The
simulations are performed long enough to allow the system to reach a
nonequilibrium steady state in which the local velocity is a
constant along the chain. To obtain a steady state, the total
integration is typically $10^{7}$ time units. We have checked that
this is sufficient for the system to reach a steady state since the
local velocity is independent of the site. For simplicity, we set
$m=\gamma=U_{0}=a=1$.

\section {Numerical results and discussion}

\begin{figure}[htbp]
  \begin{center}\includegraphics[width=7cm]{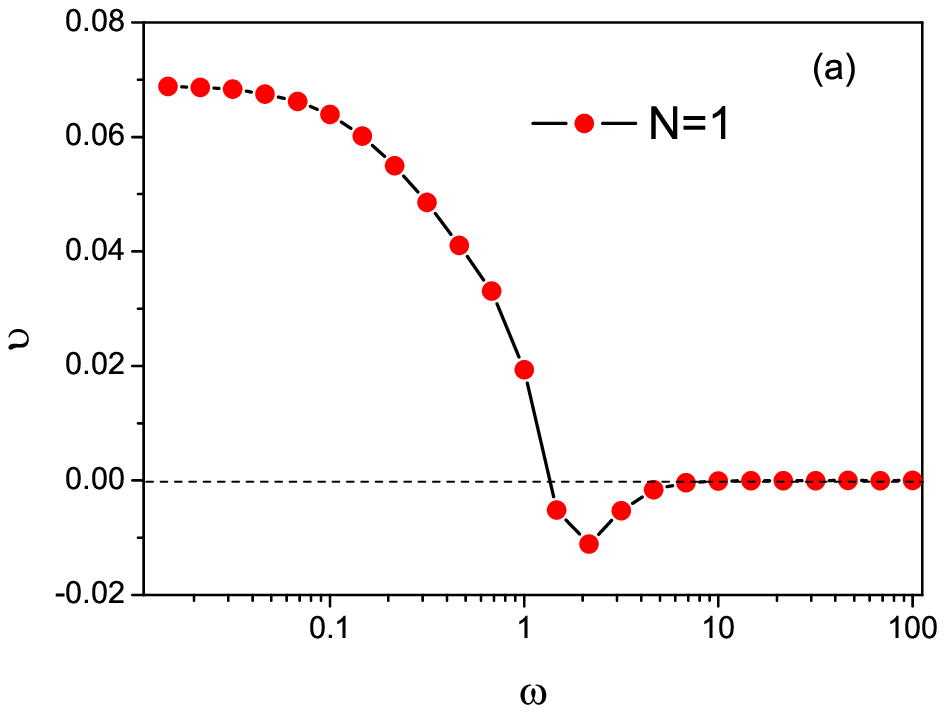}
  \includegraphics[width=7cm]{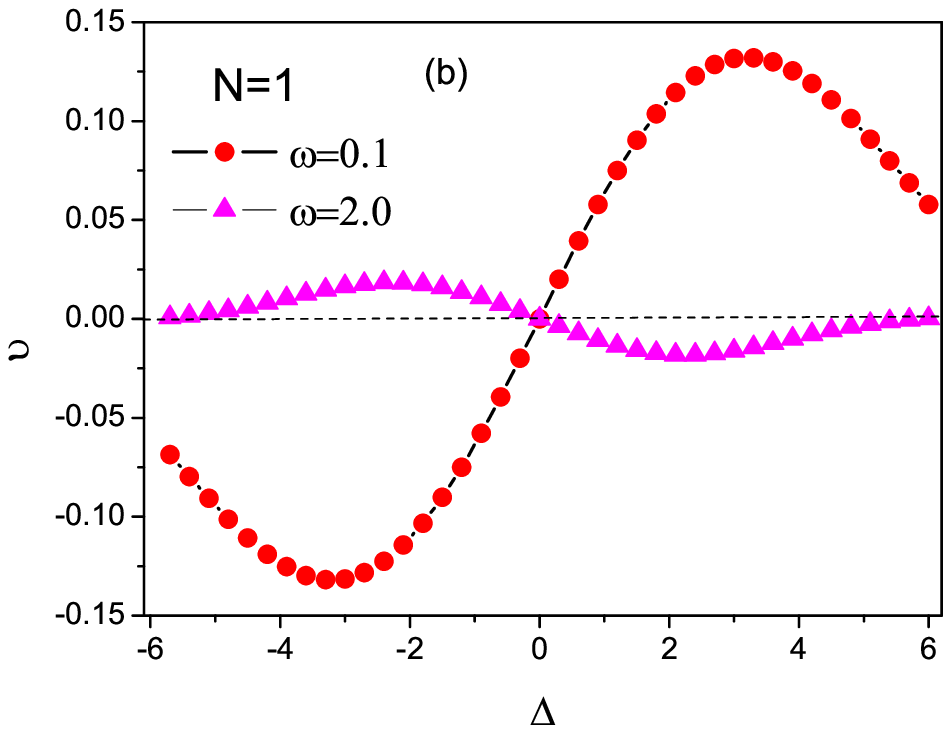}
  \caption{(Color online) Transport of overdamped Brownian particles for $N=1$.  (a) Average velocity $\upsilon$ as a function of the driving frequency $\omega$;
  (b) Average velocity $\upsilon$ as a function of the asymmetry parameter $\Delta $ of the on-site potential for $\omega=0.1$ and $\omega=2.0$.
  The other parameters are $A_{0}=2.0$, $\Delta=1.0$, and $D=0.5$.}\label{1}
\end{center}
\end{figure}

\indent Figure 3 (a) shows the average velocity as a function of the
driving frequency for overdamped case and $N=1$.  The one-particle
stochastic ratchets and stochastic resonance \cite{b2,b3} has been
extensively studied.  In the adiabatic limit $\omega\rightarrow 0$,
the ac driving force can be expressed by two opposite forces $A_{0}$
and $-A_{0}$. The particles get enough time to cross both side from
the minima of the potential. It is easier for particles to move
toward the gentler slope side than toward the steeper side, so the
average velocity is positive. As the frequency $\omega$ increases,
due to the high frequency, the particles in one period have
sufficient time to diffuse across the steeper side of the well, and
therefore it leads to a negative average velocity. When the ac
driving forces oscillate very fast $\omega\rightarrow \infty$, the
particles will experience a time average constant force
$F=\int_{0}^{\frac{2\pi}{\omega}}F(t)dt=0$, so the average velocity
goes to zero. At some intermediate values of $\omega$, the average
velocity crosses zero and subsequently reverse its direction. Figure
3 (b) depicts the average velocity as a function of the asymmetry
parameter $\Delta$ for different frequencies. For the low frequency
($\omega=0.1$), the average velocity is negative for $\Delta<0$,
zero at $\Delta=0,$ and positive for $\Delta>0$. However, for the
medium frequency ($\omega=2.0$), one can obtain the opposite
velocity, negative for $\Delta>0$ and positive for $\Delta<0$.
Moreover, for both cases, there is an optimal value of $\Delta$ at
which the velocity takes its extremal value. When $\Delta\rightarrow
0$, the system is absolutely symmetric and directed transport
disappears. When $\Delta\rightarrow \infty$, the asymmetric
potential described in Eq. (7) reduces to symmetric one with higher
barriers, $U(x)=-\frac{U_{0}}{4}\Delta \sin(2x)$, resulting in zero
velocity.

\begin{figure}[htbp]
  \begin{center}\includegraphics[width=7cm]{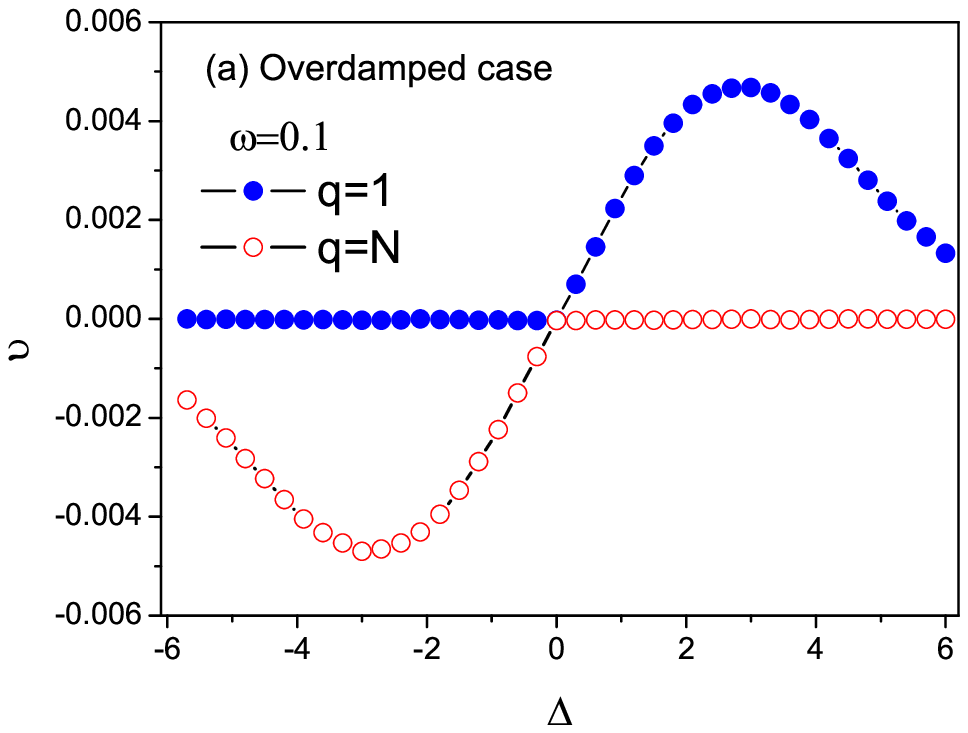}
  \includegraphics[width=7cm]{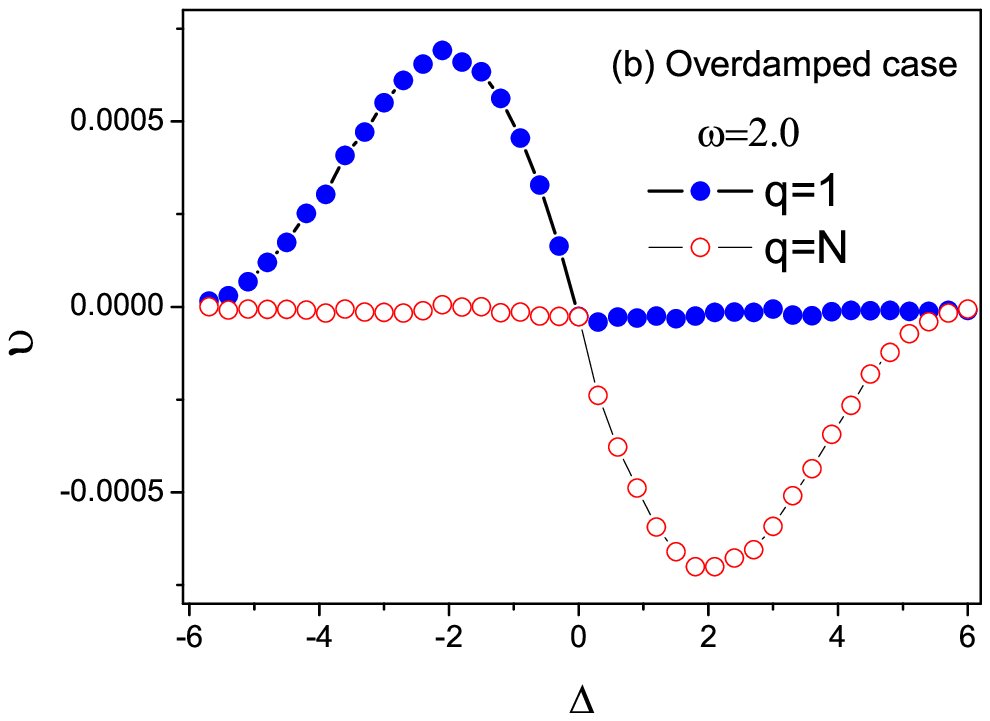}
  \includegraphics[width=7cm]{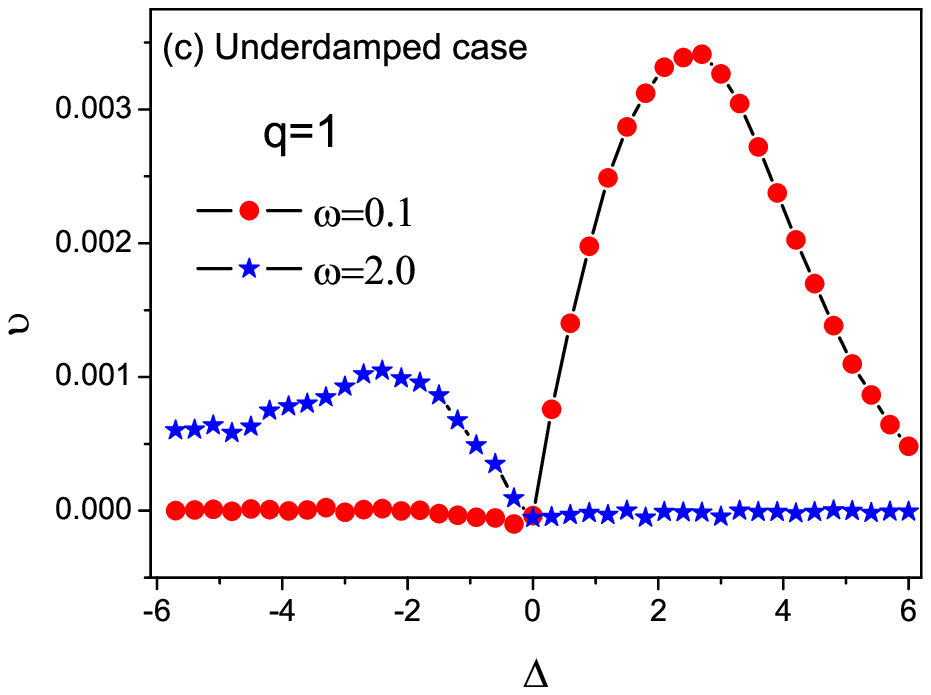}
  \caption{(Color online) Average velocity $\upsilon$ as a function of the asymmetry parameter $\Delta $ for $N>1$.
  (a) Overdamped case: $\omega=0.1$; (b) Overdamped case: $\omega=2.0$; (c) Underdamped case: $\omega=0.1$ and $\omega=2.0$.
  The other parameters are $k=1.0$, $A_{0}=2.0$, $N=11$, and $D=0.5$.}\label{1}
\end{center}
\end{figure}

\indent Now we will focus on finding how the interaction between the
particles affects the directed transport for $N>1$. Figure 4 (a)
depicts the dependence of the average velocity on the asymmetric
parameter $\Delta$ for the low frequency ($\omega=0.1$). For case A
($q=1$), the ac force acting on the left end particle, the average
velocity is positive for $\Delta>0$, it is easy for the particles to
move from the left to the right and the system is a particle
conductor. However, in the region $\Delta<0$, the average velocity
is zero, the particle can not move from the right to the left  and
the system behaves as a particle insulator. For case B ($q=N$), the
ac force acting on the right end particle, the transport from the
left to the right is prohibited and the transport from the right to
the left is permitted. The average velocity as a function of the
asymmetry parameter $\Delta$  for the medium frequency
($\omega=2.0$) is shown in Fig. 4 (b). It is found that the
unidirectional transport will also occur, but the direction of the
velocity is opposite to that in Fig. 4 (a).  From Figs. 4 (a) and 4
(b), we can conclude that the transport from the free end to the
 ac acting end is prohibited for low frequencies, while the
 transport from the ac acting end to the free end is cut off for
 medium frequencies. Therefore, the system behaves as a particle
 diode.  We also found that the similar phenomena will occur for
 underdamped case shown in Fig. 4 (c).

\begin{figure}[htbp]
  \begin{center}\includegraphics[width=7cm]{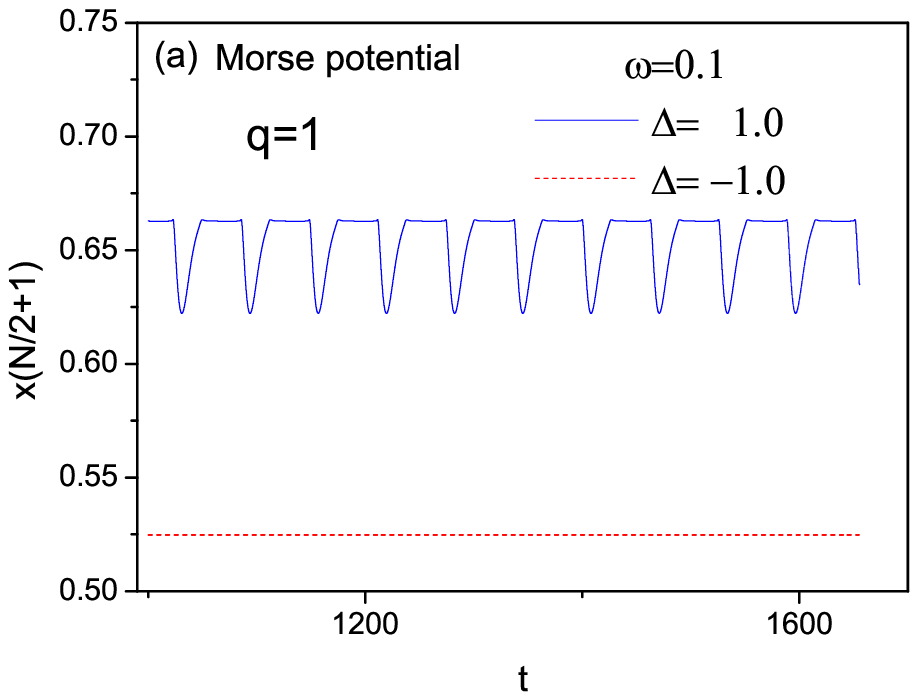}
  \includegraphics[width=7cm]{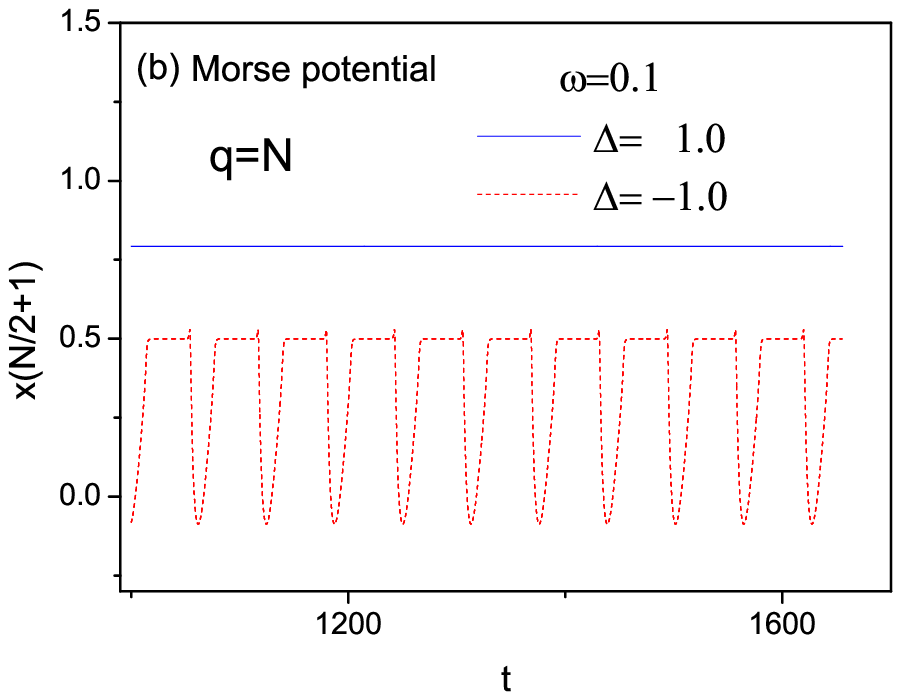}
  \includegraphics[width=7cm]{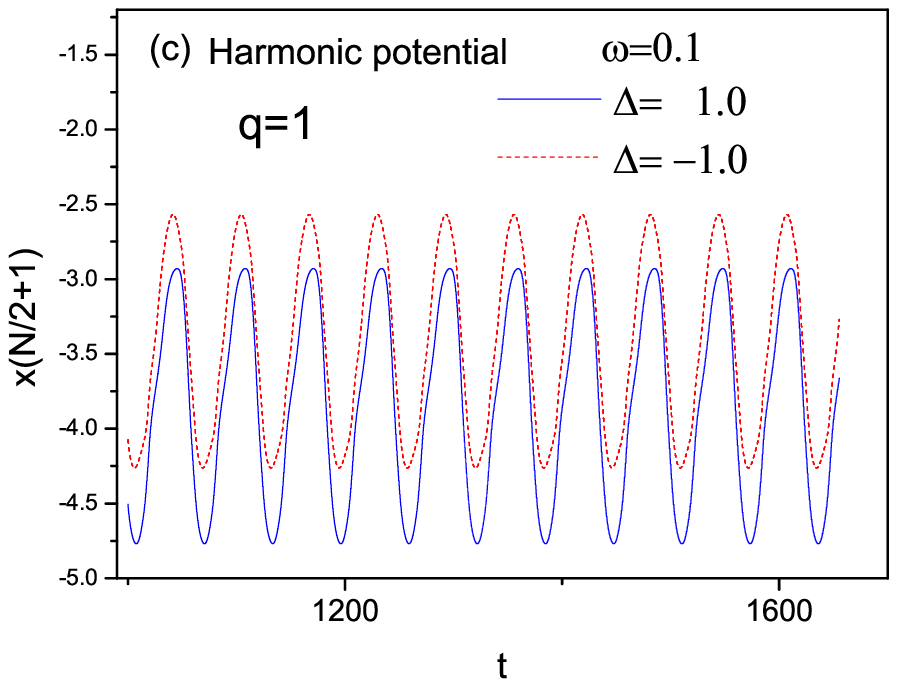}
  \caption{(Color online) Time dependent of the position of the central particle with zero noises ($D=0.0$) for overdamped case.
  (a) Morse potential and $q=1$; (b) Morse potential and $q=N$;
  (c) Harmonic potential and $q=1$. The other parameters are  $N=11$, $\omega=0.1$, $A_{0}=2.0$, and $k=1.0$. }\label{1}
\end{center}
\end{figure}

\indent Note that the similar effects of the particle diode occur
for the Lennard-Jones potential
$V(x)=4k(\frac{1}{x^{12}}-2\frac{1}{x^{6}})$, where the repulsive
force is dominated. However, the effects of the particle diode will
disappear when the interaction potential is symmetric, for example,
the Harmonic potential $V(x)=\frac{1}{2}kx^{2}$. In order to explain
the unipolarity of the transport, we investigate the motion of the
central particle with zero noises for low frequencies. For Morse
potential and case A ($q=1$), we can find that the central particle
behaves as a oscillator for $\Delta=1.0$. In this case, the driving
force acting on the first particle can be transferred to the other
particles. Therefore, directed transport will occur. However, the
central particle is static for $\Delta=-1$, the central particle can
not feel the driving force, so the transport is prohibited. For case
B ($q=N$), where the driving force is applied at the last particle,
the central particle is static for $\Delta=1.0$ and vibrates for
$\Delta=-1.0$. Figure 5 (c) shows the movement of the central
particle for Harmonic potential and case A. We can see that the
central particle vibrates for both $\Delta=1.0$ and $\Delta=-1.0$,
so the system can not behave as a particle diode. Therefore, the
asymmetry of the transport originates from the asymmetry of the
interaction potential.

\begin{figure}[htbp]
  \begin{center}\includegraphics[width=7cm]{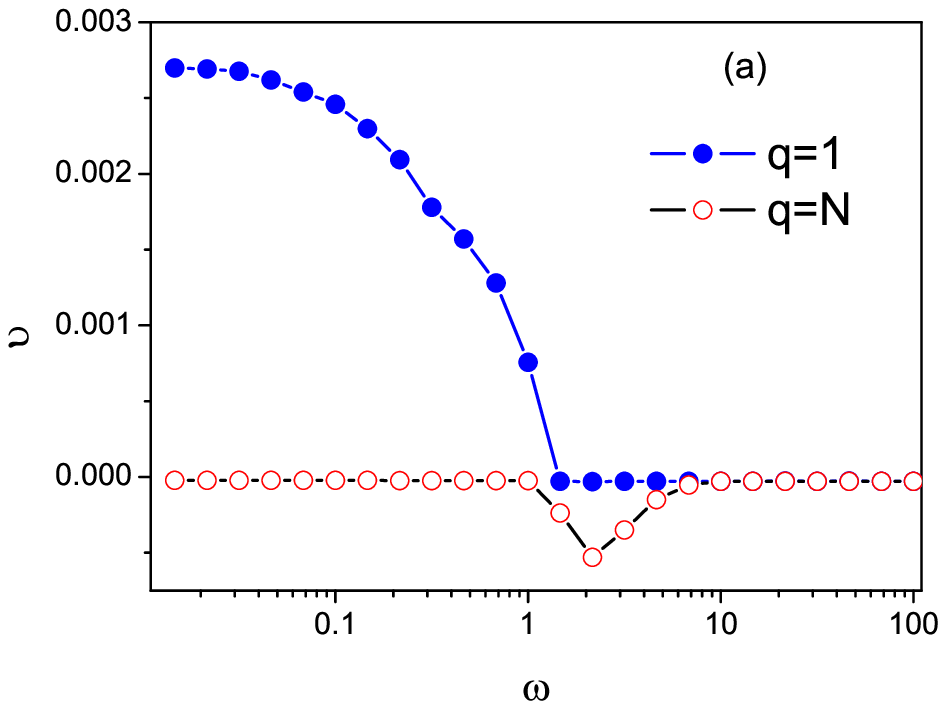}
  \includegraphics[width=7cm]{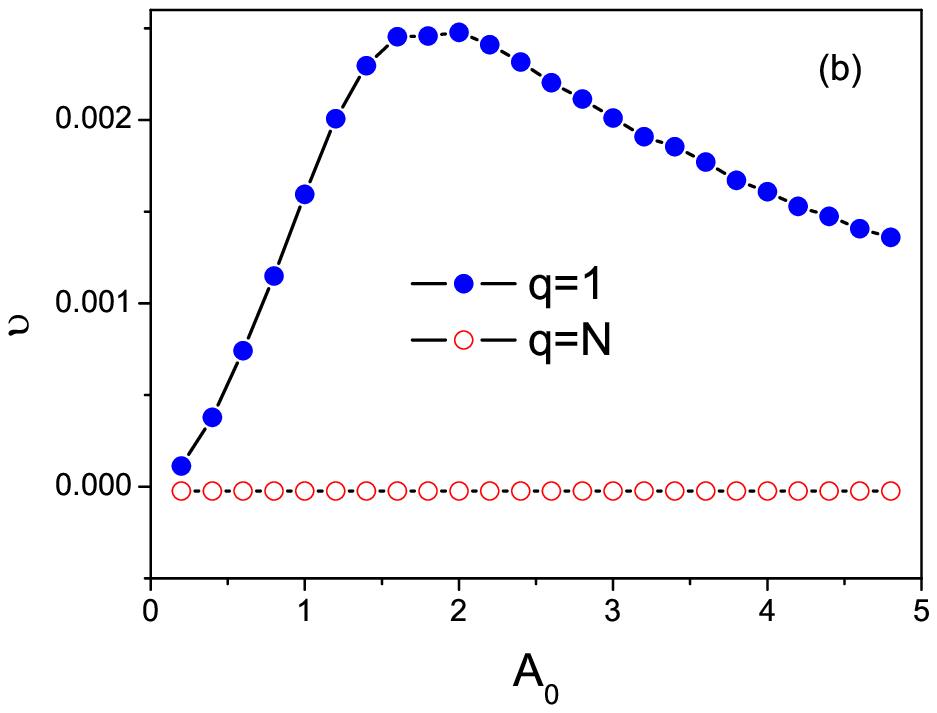}
  \includegraphics[width=7cm]{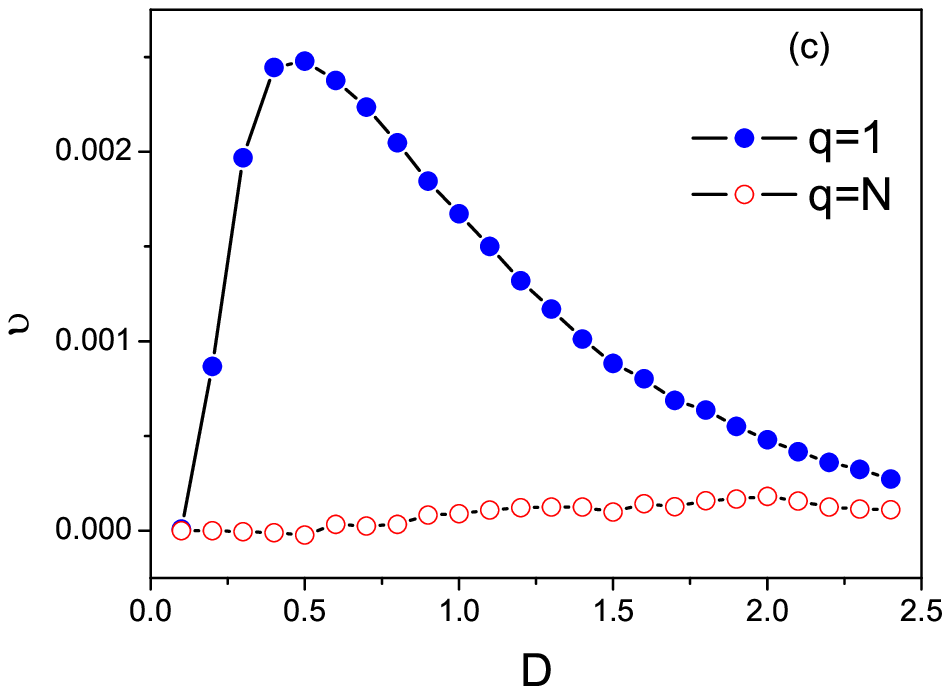}
  \caption{(Color online) Transport for overdamped case.  (a) $\upsilon$ versus $\omega$ at $A_{0}=2.0$ and $D=0.5$;
  (b) $\upsilon$ versus $A_{0}$ at $D=0.5$ and $\omega=0.1$; (c) $\upsilon$ versus $D$ at $A_{0}=2.0$ and $\omega=0.1$.
  The other parameters are $N=11$, $\Delta=1.0$, and $k=1.0$.  }\label{1}
\end{center}
\end{figure}

\indent Figure 6 (a) describes the average velocity as a function of
the driving frequency for cases A ($q=1$) and B ($q=N$) at
$\Delta=1.0$.  For low frequencies, the transport is prohibited for
case B. For medium frequencies, the average velocity is zero for
case A. The average velocity for both cases A and B is always zero
for the high frequencies.  From Figs. 6 (b) and 6(c), we can see
that the transport is prohibited for case B at $\Delta=1.0$ and
$\omega=0.1$. For case A,  there exists a value of $A_{0}$ at which
the average velocity is maximal. When $A_{0}\rightarrow 0$, the ac
driving force disappears, so the average velocity goes to zero. When
$ A_{0}\rightarrow \infty$, the driving force is very large and the
effect of the potential will disappear, the average velocity tends
to zero. Therefore, there is a peak in the curve of
$\upsilon-A_{0}$.  Similarly, the curve of $\upsilon-D$ shown in
Fig. 6 (c) for case A is observed to be bell shaped. When
$D\rightarrow 0$, the particles cannot pass across the barrier and
there is no directed current. When $D\rightarrow \infty$ so that the
noise is very large, the effect of the potential disappears and the
average velocity tends to zero, also. Therefore, one can see that
the curves demonstrate nonmonotonic behavior.

\begin{figure}[htbp]
  \begin{center}\includegraphics[width=7cm]{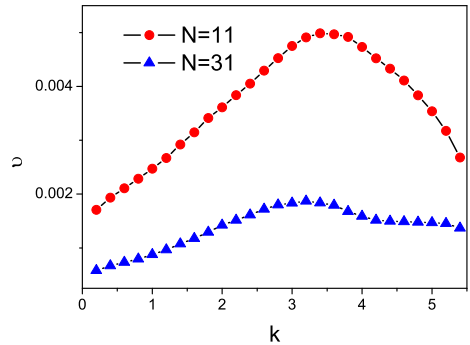}
  \caption{(Color online) Average velocity $\upsilon$ as a function of the well depth of the interaction potential $k$ for overdamped case. The other parameters are
  $\omega=0.1$, $A_{0}=2.0$,  $\Delta=1.0$, and $D=0.5$.}\label{1}
\end{center}
\end{figure}

\indent The dependence of the average velocity on the well depth of
the interaction potential is shown in Fig. 7 for overdamped case.
When $k\rightarrow 0$, the interactions between the particles
disappears,  the central particle can not experience the ac driving
force, the velocity goes to zero. When $k\rightarrow \infty$, the
particles are strongly coupled, the interacting force is very large,
the effect of the driving force will disappear, and the velocity
also tends to zero.  Therefore, there exists an optimal value of the
well depth of the interaction potential at which the average
velocity takes its maximum.

\begin{figure}[htbp]
  \begin{center}\includegraphics[width=7cm]{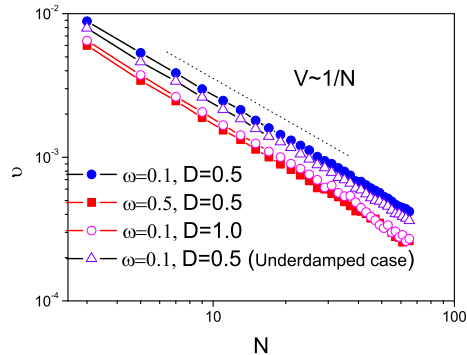}
  \caption{(Color online) Dependence of average velocity $\upsilon$ on the system size $N$ for different
  cases. For all cases the average velocity decrease in the form $\upsilon \propto N^{-1}$.
  The other parameter are $k=1.0$,  $\Delta=1.0$, and $A_{0}=2.0$. }\label{1}
\end{center}
\end{figure}
\indent We show the system size dependence of the average velocity
in Fig. 8 for different cases. It is clearly seen that the average
velocity decreases monotonically with the system size $N$ for all
cases. Remarkably, the average velocity decreases in power law
$\upsilon \propto N^{-1}$. Note that the average velocity becomes to
be independent of the system size when the driving force is applied
at every particle of the chain.

\indent Since the effect of rectification is very strong, it is
necessary to give a theoretical analysis of the observed phenomenon
by using some approximation. Here, we will focus on the small system
with $N=3$, $D=0.2$ and $\omega=0.1$ for case A. The equations of
motion for the two particles at $k=a=\gamma=1$ are
\begin{equation}\label{}
    \dot{x_{1}}=\cos(x_{1})+\frac{\Delta}{2}\cos(2x_{1})-G(x_{2}-x_{1})+\xi_{1}(t)+A_{0}\sin(\omega t),
\end{equation}
\begin{equation}\label{}
    \dot{x_{2}}=\cos(x_{2})+\frac{\Delta}{2}\cos(2x_{2})+G(x_{2}-x_{1})-G(x_{3}-x_{2})+\xi_{2}(t),
\end{equation}
\begin{equation}\label{}
    \dot{x_{3}}=\cos(x_{3})+\frac{\Delta}{2}\cos(2x_{3})+G(x_{3}-x_{2})+\xi_{3}(t),
\end{equation}
where $G(x)=-2\exp(-x)[1-\exp(-x)]$.

\indent For case of $\Delta=1.0$, when the ac force acts on the left
end,  the particle $i=1$ will jump first, and we assume that the
other particles are oscillating on average about the potential
minima. Then, Eq. (11) can be approximately written

\begin{equation}\label{}
    \dot{x_{1}}=\cos(x_{1})+\frac{\Delta}{2}\cos(2x_{1})- (\dot{x_{2}} +\dot{x_{3}})+\xi_{1}(t)+A_{0}\sin(\omega t),
\end{equation}
the term $\dot{x_{2}} +\dot{x_{3}}$ is a random and does not
dominate the transport. From this equation we can easily find that
the particle $i=1$ will move towards the right on average. The
distance between particle $i=1$ and $i=2$ is small and their
interaction becomes strong. Then the particle $i=2$ can also jump
and the ac driving force has been transferred from the particle
$i=1$ to the particle $i=2$. Therefore, Eq. (12) can
approximately become
\begin{equation}\label{}
    \dot{x_{2}}=\cos(x_{2})+\frac{\Delta}{2}\cos(2x_{2})-\dot{x_{3}}+\mu \sin(\omega t)+\xi_{2}(t),
\end{equation}
where $\mu$ is a constant. From Eq. (15), we can find that the
particle $i=2$ will also move towards the right on average.
Similarly,  the distance between the particle $i=2$ and $i=3$
becomes small, and the ac force can be transferred to the particle
$i=3$. The coupling terms prevent breaking the chain apart. Therefore,
the three particles will move together to the right (see Fig. 9
(a)) and the particles $i=2$ and $3$ behave a collective movement.

\indent However, for $\Delta=-1.0$, the average velocity based on
Eq. (14) is negative and the particle $i=1$ will move to the left on
average, the distance between the particle $i=1$ and $i=2$ becomes
large. As we know, it is easy to pull the two particles apart for
the Morse potential.  The interaction  will disappear
quickly and the particles $i=2$ and $3$ can not feel the ac driving force from
the particle $i=1$.  Therefore, Eq. (12) can be approximately written
\begin{equation}\label{}
    \dot{x_{2}}=\cos(x_{2})+\frac{\Delta}{2}\cos(2x_{2})-\dot{x_{3}}+\xi_{2}(t).
\end{equation}
From Eq. (16), we can find that the particle $i=2$ behaves thermal equilibrium and no directed transport occurs.
Therefore, the particle $i=1$ moves to the left on average and
the particles $i=2$ and $3$ have zero average velocity (see Fig. 9
(b)).  The ac driving force breaks the chain apart.  If the ac
driving can be transferred the other particles, the all particles
can move together to the same direction.  Due to the asymmetry of
the Morse potential (it is easy to pull the particles apart), the ac
driving can not be transferred to the other particle, the ac driving
particle will escape from the chain and the chain is broken apart.

\begin{figure}[htbp]
  \begin{center}\includegraphics[width=7cm]{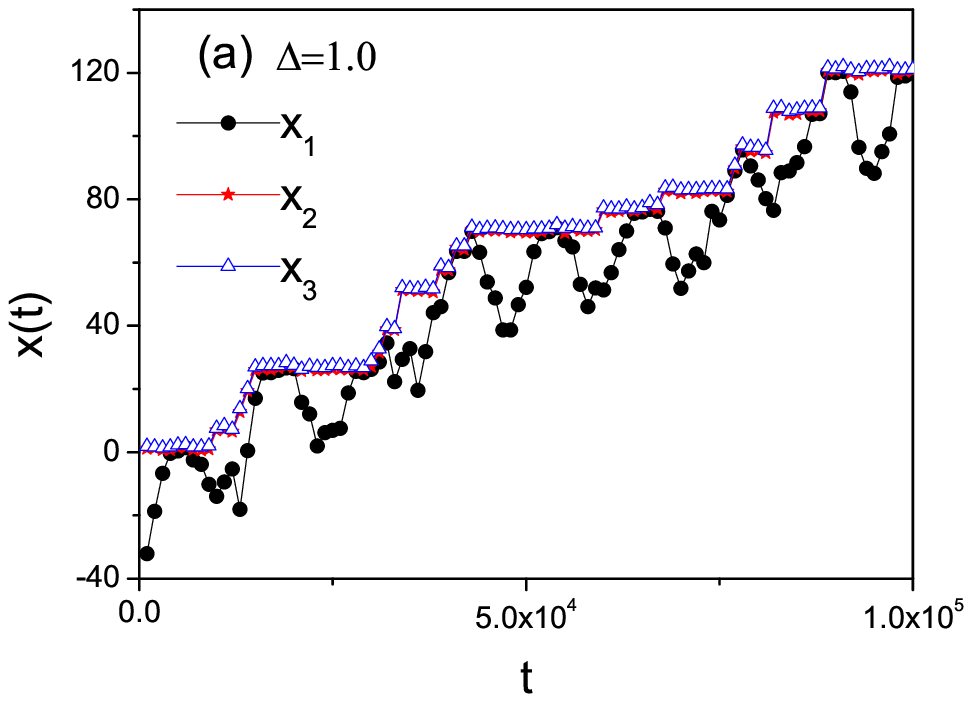}
  \includegraphics[width=7cm]{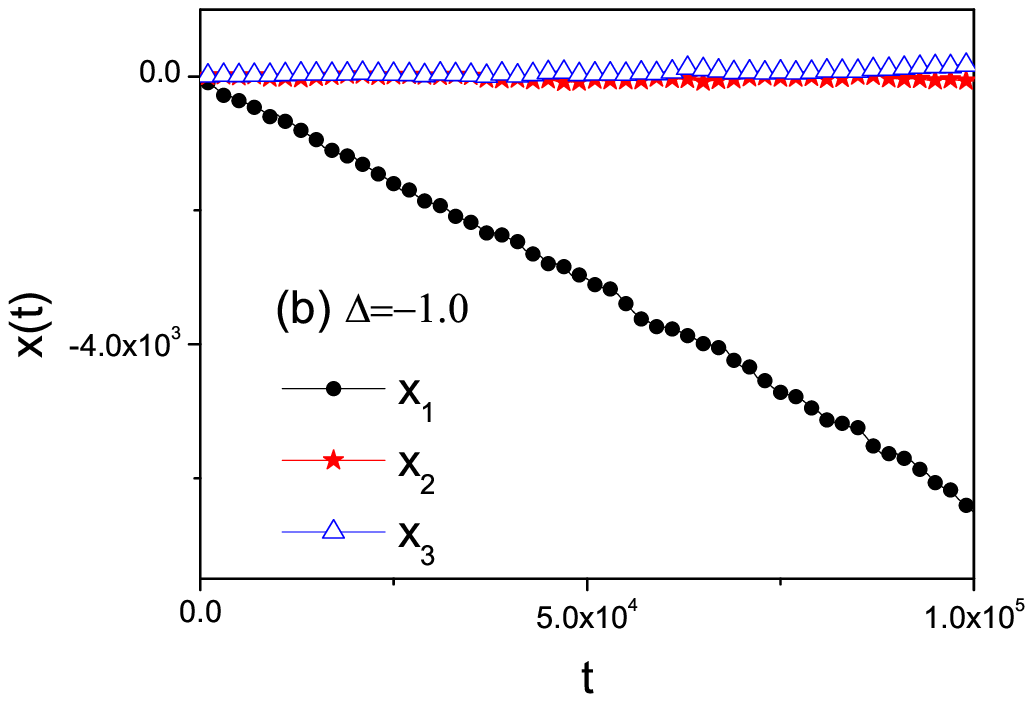}
  \caption{(Color online) The snapshots of the particles $i=1$, $2$, and $3$ for different cases.  (a) $\Delta=1.0$;
  (b) $\Delta=-1.0$. The other parameters are $k=1$, $a=1$, $A_{0}=2.0$, $\omega=0.1$, and $D=0.2$. }\label{1}
\end{center}
\end{figure}

\section{Concluding Remarks}
\indent In conclusion, we study the directed transport of
interacting Brownian particles by applying an ac driving force at
one end of the chain.  From the Brownian dynamic numerical
simulations, we obtain the average velocity of the system for both
overdamped and underdamped cases.  One can observe some features not
found in the single-particle counterparts, for example, the
unipolarity of the transport.
 When the low frequency driving force is applied at the left end particle ($q=1$), the average velocity
is positive for $\Delta>0$, the particles can easily move from the left the right and the
system works as a particle conductor. However, for $\Delta<0$,
the transport is prohibited, the particle can not move from the
right to the left and the system behaves as a particle insulator.
When the low frequency driving force is applied at the right end particle ($q=N$), the
transport from the left to the right is prohibited and the transport
from the right to the left is permitted. However, the direction of
the transport for the medium frequency force is opposite to that for the low
frequency force. We can call this system with the unidirectional transport as the particle diode.
The unipolarity of the transport is caused by the asymmetry of the interaction potential. This asymmetry of the transport will
 disappear in the symmetry interaction potential, for example, Harmonic potential.
 In addition, it is also found that there exists an optimal value of the well depth $k$
at which the average velocity is maximal. The average velocity $\upsilon$
decreases monotonically with the system size $N$, $\upsilon\propto N^{-1}$.

\indent This work was supported in part by the National Natural
Science Foundation of China (Grants No. 61072029, No. 11004082, and
No. 10947166) and the Guangdong Provincial Natural Science
Foundation (Grants No. 01000025 and No. 01005249). Y. -f. H. also
acknowledges the Natural Science Foundation of Hebei Province of China (Grant No. A2011201006).

\end{document}